\documentclass[aps,prd,twocolumn,showpacs,nofootinbib]{revtex4-1}

\usepackage{amsmath}
\usepackage{amsfonts,amssymb}
\usepackage{multirow}
\usepackage[dvips]{graphicx}
\usepackage{float}
\usepackage{appendix}
\usepackage{color}
\usepackage{url}
\usepackage{subfigure}
\usepackage{footnote}

\usepackage{mathrsfs} 
\usepackage{slashed}

\def\beq{\begin{equation}}
\def\eeq{\end{equation}}
\def\bea{\begin{eqnarray}}
\def\eea{\end{eqnarray}}
\def\nn{\nonumber}
\def\nl{\nonumber\\}

\def\chic1{\chi_{c1}}

\newcommand{\vp}{\mathbf{p}}
\newcommand{\vep}{\boldsymbol{\varepsilon}}

\def\X{X(5568)}

\def\Fbgr{F_\text{bgr}}

\def\fbgr{f_\text{bgr}}
\def\fsig{f_\text{sig}}

\def \Tr {\text{Tr}\,}

\RequirePackage{xspace} \allowdisplaybreaks

\begin{document}

\title{$P$-wave coupled-channel scattering of $B_s\pi,\,B_s^*\pi,\,B\overline K,\,B^{*}\overline K$ and the puzzling $X(5568)$} 
\author{Xian-Wei Kang }\email{xianwei.kang@um.es}
\author{J.~A.~Oller }\email{oller@um.es}
\affiliation{Departamento de F{\'i}sica, Universidad de Murcia, E-30071 Murcia, Spain}

\begin{abstract}
The D0 Collaboration has recently reported a narrow peak structure in the $B_s^0\pi^\pm$ invariant-mass spectrum, that is called $X(5568)$. 
Its proposed quantum numbers $J^P=0^+/1^+$ are not determined unambiguously, and we show that one can fit the data  assuming $J^P=1^-$ with 
the same mass and width for the resonance signal. The corresponding isovector $P$-wave coupled-channel scattering involving the states 
$B_s\pi$, $B_s^*\pi$, $B \overline{K}$ and $B^*\overline K$ is studied employing leading-order Heavy Meson Chiral Perturbation Theory, which is
then unitarized making use of standard techniques from Unitary Chiral Perturbation Theory. The subtraction constants 
that appear in the unitarization process are determined by a $u$-channel crossing symmetry constraint 
or by naturalness arguments, being the numerical values obtained in both cases quite closed, and no further freedom is left in the model. 
  A crucial point is that if the  $X(5568)$ is $1^-$ and decaying into $B_s^0\pi^\pm$ then it should appear as a pole in the  $T$-matrix elements. 
 It finally turns out that no $X(5568)$ pole  is found that can be qualified as  dynamically generated 
from the isovector $P$-wave coupled-channel dynamics. This result disfavors the interpretation of the $\X$ within a ``molecular'' picture.
\end{abstract}

\maketitle

\section{Introduction}

Recently, the D0 Collaboration has reported a narrow peak in the $B_s^0\pi^\pm$ mass spectrum, denoted as $X(5568)$,
 based on their 104 fb$^{-1}$ data of $p\bar p$ collision at $\sqrt{s}=1.96$~TeV accumulated at the Fermilab Tevatron collider \cite{D0}.
 The resulting mass and width of this peak were determined as 
\bea\label{eq:pole position}
m_X&=&5567.8\pm2.9^{+0.9}_{-1.9}\,\text{MeV},\nl
\Gamma_X&=&21.9\pm6.4^{+5.0}_{-2.5}\,\text{MeV},
\eea
respectively, and the spin-parity assignment $J^P=0^+$ was used in the data analysis (with $1^+$ also discussed). 
In fact, the quantum numbers $J^P$ have not been well determined yet. 
The statistical significance was found to be $5.1\sigma$, including look-elsewhere effect and systematic uncertainties \cite{D0}. 
 The decay of $X(5568)$ to $B_s^0\pi^\pm$ suggests that it has four valence quarks with different flavors: $su\bar{b}\bar{d}$ or $sd\bar{b}\bar{u}$. 
If finally the existence of the $X(5568)$ is confirmed, this will  be the first observation of a resonance with all these quark flavors.
This new finding has triggered the interest of theorists, and many interpretations have been proposed. 
 Particularly, motivated by the aforementioned four quark components, one has models for tetraquark states
\cite{tetraquarkWangW,tetraquarkChenW,tetraquarkQiao,LiuTetra}, which typically show that their outcome is compatible  with the experimental mass.
 Through a comparison between the molecular assignment and the diquark-antidiquark scenario within 
QCD sum rules, the tetraquark state is advocated in Ref.~\cite{molecular}.
Although a four-quark ($[su][\bar b\bar d]$) state with $J^P=0^+$ was found about 150 MeV higher than the $\X$ using the
diquark-antidiquark interaction in Ref.~\cite{tetraquarkWangW}, the authors also claimed that the state is still likely
 to be identified as the $\X$ once systematic errors in the model are considered.
 This state has also been analyzed within the quark model to further investigate the tetraquark components \cite{quarkmodel-simple,quarkmodel-Dong}.
In the former reference the calculated mass for $\X$ is rather close to the experimental one, but  the author also asks in the paper for a more 
elaborated model  for the $su\bar{d}\bar{b}$ system.
The subsequent calculation in the relativized quark model \cite{quarkmodel-Dong} finds that the masses of $sq\bar b\bar q$
 tetraquark states are much higher than the one of $X(5568)$, and thus the interpretation as a 
 diquark-antidiquark state is disfavored. Later the dispute becomes more severe as LHCb fails to confirm such peak \cite{LHCb}
 using its 3 fb$^{-1}$ $pp$ collision data at $\sqrt{s}=7$ and 8 TeV. In fact, several theoretical investigations e.g. Refs.~\cite{Swanson,FKGuo} point out
that none of the models, including the threshold and cusp effects, molecular and tetraquark ones, give a satisfactory description of the $X(5568)$.

Already Ref.~\cite{Swanson} discussed the possibility of the 
$\X$ being generated from the $S$-wave coupled-channel transition of $B_s \pi\to B \overline K$, and performed 
a corresponding calculation within the nonrelativistic quark model. The authors concluded that  this scenario is unlikely. 
This idea has been also implemented in the work of Ref.~\cite{Oset5568},
 where the $S$-wave potential between $B_s\pi$ and $B\overline K$  was calculated in the leading-order
Heavy Meson Chiral Perturbation Theory (HMChPT), and then unitarized with the method of Ref.~\cite{NPA620}. The free parameter, a subtraction
constant  or a three-momentum cutoff, is fitted to the experimental data on the $B_s^0\pi^\pm$ invariant-mass spectrum. 
It turns out that an  unnaturally large cutoff, much larger than 1 GeV, is unavoidably 
needed to reproduce the mass of the resonance (with the resulting width in agreement with experiment).
 In this sense, the $X(5568)$ is hard to be assigned as a dynamically generated resonance by the $S$-wave $B_s \pi\to B\overline K$  scattering, 
 disfavoring the  molecular interpretation.

In fact, the quantum numbers $J^P$ for the resonance have not been  determined unambiguously yet.
 The assignments $0^+$ ($\X\to B_s^0\pi^\pm$) or 
$1^+$ ($\X\to B_s^{*}\pi^\pm$ followed by $B_s^*\to B_s^0\gamma$)  from Ref.~\cite{D0}
 are purely based on fitting the resonance signal in the $B_s^0\pi^\pm$ event distribution with an $S$-wave Breit-Wigner (BW) function. 
  We have also performed a purely phenomenological fit to the data of Ref.~\cite{D0} 
in terms of a $P$-wave BW parameterization of the signal function with the same mass and width as in Eq.~\eqref{eq:pole position}, 
 and show that a reproduction nearly as good as the one assuming an $S$-wave BW results.  In other terms,  the current data is not capable
 of unambiguously pinning down the quantum numbers (and in particular the intrinsic parity) of the $\X$.  
Another point that we want to stress here is that the mass threshold of $B_s^{*}\pi^+$
 is below the pole position of $X(5568)$ only by around 10 MeV, closer indeed to the 
 $X(5568)$ mass than the $B_s^0\pi^+$ threshold. Thus, from a ``molecular'' perspective  for the generation of the $\X$  it is suggestive  
to look for a coupled-channel-scattering scenario involving simultaneously $B_s\pi$ and $B_s^*\pi$. 
 However, in order to conserve angular momentum and parity it is necessary at least a $P$-wave to couple both channels. 
 It is typically considered that the $P$-wave scattering between $B_s\pi$ and $B_s^*\pi$ is small in the $\X$ region 
because their three-momenta are small. However, 
the off-diagonal transitions involving the heavier $B\overline{K}$ and $B^*\overline{K}$ states have much larger {\it unphysical} three-momenta, 
determined by analytical continuation. 
 This motivates our current study in coupled channels within HMChPT to study the isovector $P$-wave scattering between the channels
 $B_s\pi$, $B_s^*\pi$, $B\overline{K}$ and $B^*\overline{K}$. 
 We also discuss that the subtraction constant per channel appearing in the 
unitarization process can be pinned  down reliablly by imposing an $u$-channel crossing symmetry constraint, similarly as done 
in Refs.~\cite{alarcon.210616.1,alarcon.210616.2}. The numerical values obtained in this way are also quite closed to the natural values 
 that stem from a three-momentum cut-off around 1~GeV \cite{OllerMeissnerPLB}.
   We  conclude from our research that an explanation of the $\X$ as dynamically 
generated resonance in $P$-wave scattering is not plausible.

This work is organized as follows. In Sec.~\ref{P-wave scenario}, 
we discuss  the reproduction of the  D0 data employing a  $P$-wave BW parameterization for the signal function. 
Next we move to HMChPT in order to calculate the leading-order tree-level amplitudes projected in $P$-wave (Sec.~\ref{sec:potential}) 
and then its unitarization and following results are  discussed  in Sec.~\ref{sec:results}.
 At last, we conclude in Sec.~\ref{sec:conclusion}.

\section{Interpretation of D0 data with  $B_s^0\pi^\pm$ in $P$ wave}
\label{P-wave scenario}

The $B_s^*\pi$ mass threshold is very close to the pole position of the $X(5568)$, as mentioned in the Introduction.
 The coupling of $B_s^*\pi$ to $B_s^0\pi$ requires at least a $P$-wave, 
because  $S$-wave is not allowed due to angular momentum and parity conservation. 
These observations encourage us to reexamine the interpretation of the experimental data with $B_s^0\pi^\pm$ in $P$ wave and not 
in $S$ wave, as originally performed by the D0 Collaboration \cite{D0}. In this form the vector and pseudoscalar heavy mesons could scatter 
between each other within $P$-wave coupled-channel scattering (as analyzed in detail below in Sec.~\ref{sec:potential} and \ref{sec:results}). 

Before proceeding to the $P$-wave case, we review the formalism used by the  D0 Collaboration \cite{D0} in fitting the 
invariant mass distribution of $B_s^0\pi^\pm$ assuming $S$-wave.
 The original full data contains both the background and the resonance signal.
 The extraction of the latter was done by fitting data without and with the ``cone cut'', 
as introduced in Ref.~\cite{D0} in order to enhance the resonance signal. 
 Both  background components, that result with or without this cut, can be well parameterized as \cite{D0talk}
\begin{equation}\label{eq:bkg}
\Fbgr(m)=(c_0+c_2 m^2+c_3 m^3+ c_4 m^4) \exp(c_5+c_6 m+c_7 m^2)~,
\end{equation}
where $m\equiv m(B_s^0\pi^\pm)$ is the invariant mass of $B_s^0\pi^\pm$ (we use this shorthand notation in the following, if not otherwise stated). 
This parameterization reproduces perfectly well the background in all cases \cite{D0talk,D0}.

 A BW parameterization for a resonance of mass $M_X$ and width $\Gamma_X$ near the two-body threshold
 with orbital angular momentum $L$ can be  written as 
\begin{equation}\label{eq:BW}
BW_L(m)\propto \frac{M_X^2\Gamma_X(m)}{(M_X-m)^2+M^2\Gamma_X^2(m)}~.
\end{equation}
 This is written in terms of the mass-dependent width, which  reads
\begin{equation}
\Gamma(m)=\Gamma_X \left(\frac{k(m)}{k(M_X)}\right)^{2L+1}~,
\end{equation}
and $k(m)$ is the magnitude of the center-of-mass (CM) three-momentum of $B_s^0\pi^\pm$
 with  invariant mass $m$, see  Eq.~\eqref{eq:kappa} below for an explicit expression.
 As a result, the $P$-wave BW parameterization corresponds to replacing the factor $k(m)/k(M_X)$ in Ref.~\cite{D0}
by $(k(m)/k(M_X))^3$; i.e., the threshold behavior for the transition amplitude is implemented, 
see also the review ``Resonances'' in PDG \cite{pdg}.

\begin{figure}[htbp]
\begin{center}
\includegraphics[height=120mm,clip]{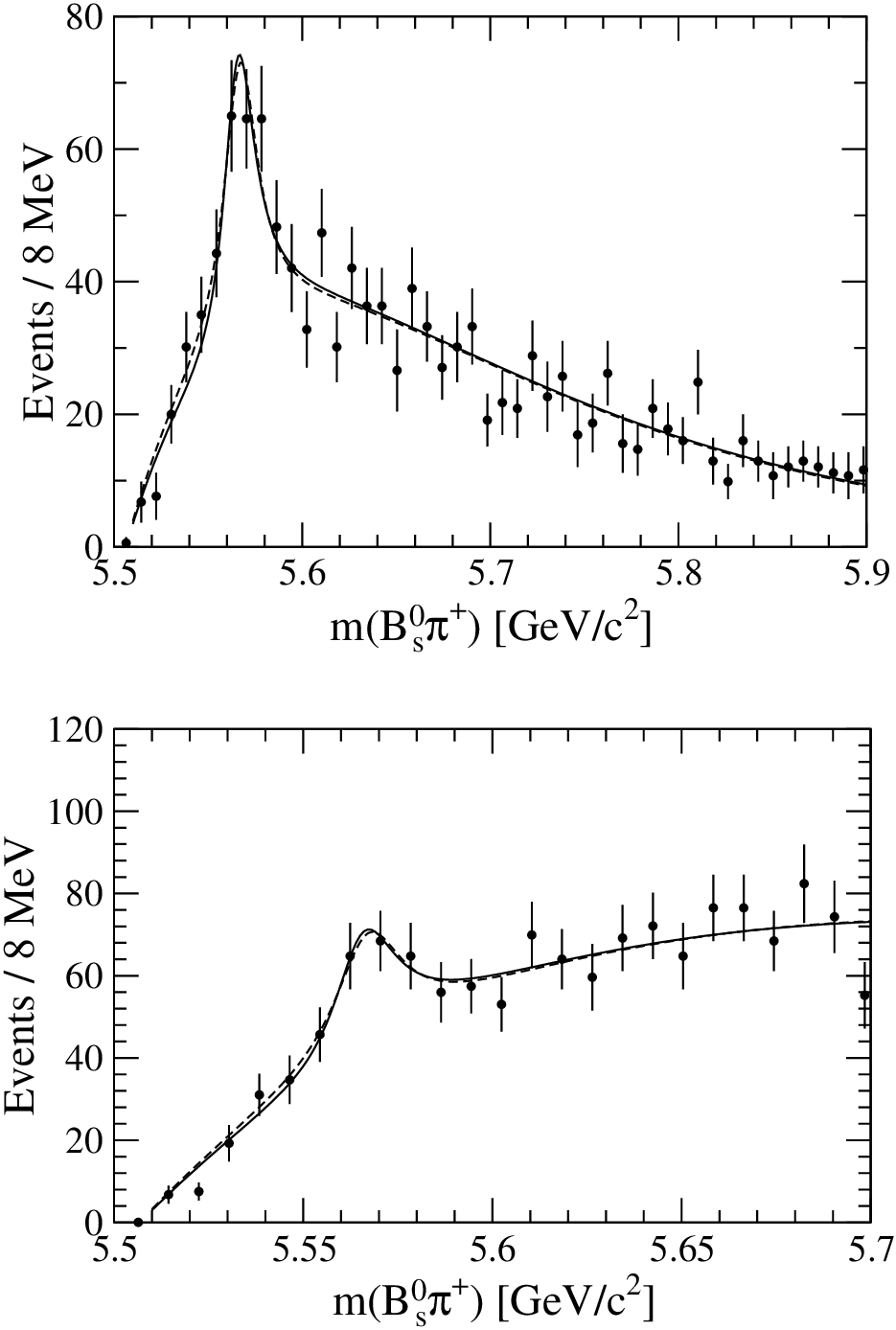}
\caption{Invariant mass distribution  for $B_s^0\pi^\pm$  with the data  taken from Ref.~\cite{D0}.   
The upper (lower) panel corresponds to the case with (without) the ``cone cut'' \cite{D0}.
 The solid (dashed) lines represent our results with the mass and width 
of the resonance given by Eq.~\eqref{eq:pole position} for the $P$-($S$-)wave BW parameterizations of the signal function.}
\label{fig:full data}
\end{center}
\end{figure}

 Analogously to Ref.~\cite{D0} we first fit the full data with the function
\begin{equation}
\label{eq:F(m)}
F(m)=\fsig\times BW_0(m)+\fbgr\times \Fbgr(m)~,
\end{equation}
where the resonance signal is parameterized in terms of an  $S$-wave BW times a normalization constant $\fsig$. 
The background is given in terms of  $\Fbgr(m)$, cf. Eq.~\eqref{eq:bkg}, times the constant $\fbgr$. 
 Both  $\fsig$ and $\fbgr$ are fitted to data, while  $M_X$ and $\Gamma_X$ are fixed according to their experimental
 numbers in Eq.~\eqref{eq:pole position} obtained by the D0 Collaboration. 
The resulting curves are given by the dashed lines in Fig.~\ref{fig:full data} (the upper panel corresponds to the case with the ``cone cut'' and the 
 bottom one without this cut) and closely reproduce data.

However, the parameterization of the signal function by an $S$- or $P$-wave BW cannot be really ascertained by the present experimental data 
\cite{D0}. This can be easily seen by directly replacing $BW_0(m)$ in Eq.~\eqref{eq:F(m)} by $BW_1(m)$ (with the same mass and width)
 and refitting $\fsig$ and $\fbgr$. The change for these parameters is small, being around a 5\% for the former and a 2\% for the latter. 
 The result of this fit is shown by the solid lines in Fig.~\ref{fig:full data}, which can hardly be distinguished from the dashed
 lines corresponding to the $S$-wave BW parameterization of the signal function. Because of the close proximity between the $S$- and $P$-wave 
BW fits  and their good reproduction of data, 
we skip distinguishing between partial and total decay widths of the $X(5568)$ in Eq.~\eqref{eq:BW}. Notice that  for a $1^-$  resonance 
the $B_s^{*}\pi^+$ channel is also open since $M_X$ is slightly larger (around 10~MeV) than its threshold.

  Thus, one cannot exclude  a $P$-wave scenario for the invariant mass distribution of $B_s^0\pi^\pm$ from the decay of the $X(5568)$ resonance,
with the present data determined by the D0  Collaboration \cite{D0}. 
As remarked in Ref.~\cite{LiuXH}, an angular distribution analysis should be performed in order to further constraint 
 the quantum numbers of the $X(5568)$ and check any model on this resonance. 
 In the rest of this work we are going to consider the assignment $J^P=1^-$ for the $X(5568)$, a vector state, and 
explore its possible dynamical generation from coupled-channel scattering of 
 $B_s^0\pi^+$,  $B_s^*\pi^+$, $B^+\overline K^0$ and $B^{*+}\overline K^0$.

 \section{Coupled-channel potential}
\label{sec:potential}

Assuming that the $X(5568)$ is an $1^-$ state, $B_s^0\pi^+$ scattering occurs in $P$-wave, in which case, the transition between $B_s^0\pi^+$ (1), $B_s^*\pi^+$ (2), 
 $B^+\overline K^0$ (3) and  $B^{*+}\overline K^0$ (4) are allowed. In this list the channels have been labeled in increasing order of
 their mass thresholds: 5506.36 MeV, 5554.97 MeV, 5772.29 MeV, 5818.10 MeV, respectively. 
Such a coupled-channel system can be treated in Heavy Meson Chiral Perturbation Theory (HMChPT) \cite{Wise,Yan,Donoghue}.
Here, we closely follow  Wise's convention in Ref.~\cite{Wise}. 
The interactions of the heavy mesons ($B^{(*)}$ or $B_s^{(*)}$) and the pseudo-Goldstone bosons ($\pi,\,K,\,\eta$) are described at leading order in 
HMChPT by the Lagrangian
\bea \label{eq:LagWise}
\mathscr{L}&=&\frac{i}{2} \Tr \overline H_a  H_b  v^\mu  (\xi^\dagger \partial_\mu\xi + \xi\partial_\mu\xi^\dagger)_{ba}\nl 
&&+\frac{ig}{2} \Tr \overline H_a  H_b \gamma^\mu \gamma_5 (\xi^\dagger \partial_\mu\xi - \xi\partial_\mu\xi^\dagger)_{ba}\, ,
\eea
with the $SU(3)$ subscripts $a$ and $b$ running from 1 to 3. 
The quantity $v$ is the four-velocity of the heavy meson, whose four-momentum can be parameterized as 
\beq
p= m \,v \, + \, q, 
\eeq
with $q$ a small residual four-momentum. The $H_a$ fields are defined as
\bea
H_a&=&\frac{1+ \slashed v }{2}\left(P_{a\mu}^*\gamma^\mu-P_a\gamma_5\right) \,,\nl
\overline H_a &=& \gamma^0 H_a^\dagger \gamma^0 \,,
\eea
with the heavy meson fields containing the heavy $b$ quark: 
\bea
(P_1, P_2, P_3)&=&(B^-, \overline B^0, \overline B_s^0)\,,\nl
(P_1^*, P_2^*, P_3^*)&=&(B^{*-}, \overline B^{*0}, \overline B_s^{*})~,
\eea
and  $P$ ($P^*$) denotes a heavy pseudoscalar (vector) meson. 
The symbol $\Tr$ in Eq.~\eqref{eq:LagWise} indicates the  trace taken over the Dirac matrices. 
The light pseudoscalar mesons are contained in the $3\times 3$ matrix $\xi$, defined as
\beq
\xi=\exp\left( \frac{i\phi}{\sqrt{2}f_\pi}\right)~,
\eeq
where
\beq
f_\pi\approx 92.4\,\,\text{MeV}
\eeq
is the (leading-order) pion decay constant and 
\beq
\phi=\begin{pmatrix} 
\frac{\pi^0}{\sqrt{2}} + \frac{\eta}{\sqrt{6}}  & \pi^+                                                  & K^+ \nl
\pi^-                                           &-\frac{\pi^0}{\sqrt{2}} + \frac{\eta}{\sqrt{6}}         & K^0 \nl
K^-                                             & \overline K^0                                          & -\sqrt{\frac{2}{3}}\eta                                              
 \end{pmatrix}\, .
\eeq
 In the original reference \cite{Wise},
 another convention on the pion decay constant, $f=\sqrt{2}f_\pi\approx 130$ MeV, was used.
Note also that $v_\mu P^{*\mu}=0$ due to the Lorenz condition.

 Explicitly, from the Lagrangian in Eq.~\eqref{eq:LagWise} we obtain the following terms relevant for $B_s^0\pi^+$ plus coupled-channel scattering 
\bea \label{eq:Lagexplicit}
\mathscr{L}&=&- \frac{\sqrt{2}g}{f_\pi}\left( P_a^{*\dagger\alpha}\, P_b+P_a^\dagger P_b^{*\alpha}\right) (\partial_\alpha \phi)_{ba}\, \nl 
           && -\frac{i\sqrt{2}g}{f_\pi}\, \epsilon^{\mu\nu\sigma\alpha}\, P_{a\mu}^{*\dagger}\, P_{b\nu}^*\, v_\sigma\, (\partial_\alpha \phi)_{ba} \nl
           && +\frac{i}{2f_\pi^2}\, P_{a\mu}^{*\dagger}\, P_{b}^{*\mu}\,v^\alpha\,  (\phi\,\partial_\alpha\phi - \partial_\alpha\phi\,\phi)_{ba}\nl
           && -\frac{i}{2f_\pi^2}\, P_a^{\dagger}\, P_b\, v^\alpha\,  (\phi\,\partial_\alpha\phi - \partial_\alpha\phi\,\phi)_{ba}+\ldots
\eea
from which the interaction vertices can be read off easily.  The ellipsis in the previous equation indicates terms that also contribute to ${\cal L}$ 
but are not of interest here.
 Each term in Eq.~\eqref{eq:Lagexplicit} from the the first to the fourth line gives rise to a vertex involving 
  $P^*P\phi,\, P^* P^*\phi,\, P^* P^*\phi\phi,\, PP\phi\phi$, respectively.
 For our present purposes each  $P$ ($P^*$) denotes a heavy pseudoscalar (vector) belonging to the $B$ ($B^*$) family, 
and every $\phi$ contributes with a pseudo-Goldstone boson ($\pi,\, K,\, \eta$). 
The fact that $B^*$ cannot decay to $B\pi$ ($m_{B^*}< m_B + m_\pi$) complicates the determination of $g$ by experiment in the $B$ sector. 
However, it has been studied extensively within lattice QCD. The most recent determination
 can be found in Ref.~\cite{Flynn}, where $g_{B^*B\pi}=0.56\pm 0.08$ is reported,
where we have added the statistical and systematic errors in quadrature.
 Previous results from lattice QCD calculations are also organized and compared in this reference, and they agree within uncertainties.
 Under the heavy quark flavor symmetry, one has also that $g=g_{D^*D\pi}$, which can be rather well determined from the experimental decay width of $D^{*+}\to D^0\pi^+$. 
The result is $g=g_{D^*D\pi}=0.58 \pm 0.07$, with the error obtained from the uncertainty of the $D^{*+}$ width \cite{KangBl4}. 
This is in a perfect agreement with the lattice QCD results. In the numerical calculations we use the value determined from phenomenology,
 as also done in Ref.~\cite{KangBl4}.

\begin{figure}[htbp]
\begin{center}
\vglue 0.5cm
\includegraphics[height=85mm,clip]{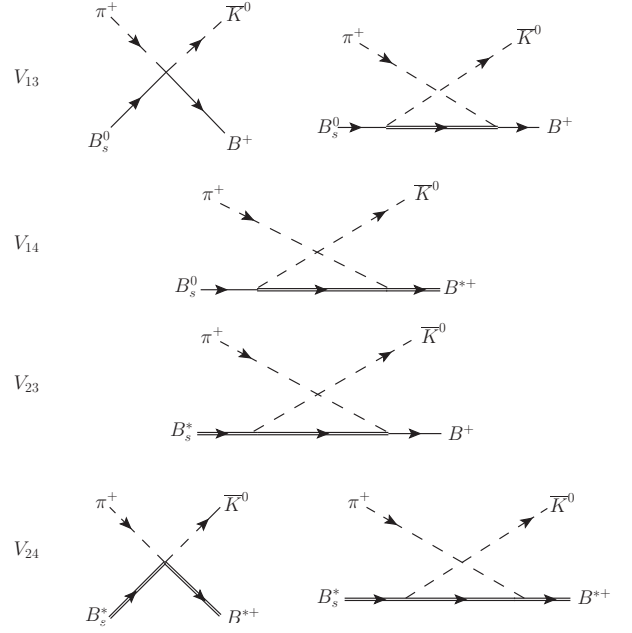}
\caption{The leading-order Feynman diagrams in HMChPT contributing to the $4\times 4$ potential matrix $V_{ij}$. 
 The double (single) lines represent heavy vectors (pseudoscalars)
 and the dashed lines denote the light pseudoscalars $\pi^+$ or $\overline K^0$.} 
\label{fig:diagram}
\end{center}
\end{figure}

Based on the above Lagrangian, we can list out and calculate the Feynman diagrams at the leading order, which are drawn in Fig.~\ref{fig:diagram}. 
Here the heavy vector (pseudoscalar) mesons are indicated by double (single) solid lines and the light pseudoscalar mesons ($\pi^+$ and $\overline{K}^0$) 
by  dashed lines. 
From the first to fourth line, they contribute to
$V_{13}\equiv V_{B_s^0\pi^+\to B^+ \overline K^0}$,\, $V_{14}\equiv V_{B_s^0\pi^+\to B^{*+} \overline K^0}$\, $V_{23}\equiv V_{B_s^*\pi^+\to B^+ \overline K^0}$ and
$V_{24}\equiv V_{B_s^*\pi^+\to B^{*+} \overline K^0}$, respectively. 
In $V_{13}$ and $V_{24}$, there are two diagrams, labeled by the subscripts 1 and 2, from left to right. 
All other contributions vanish in the $4\times4$ potential matrix and, in particular, 
 there are neither $s$-channel nor $t$-channel exchange diagrams at leading order. 
 After the calculation, we obtain the potentials in the rest frame of heavy particles (i.e. $v=(1, 0, 0, 0)$):
\bea \label{eq:potential}
V_{13,1}&=&\frac{\sqrt{m_{B} m_{B_s}}}{2f_\pi^2} (E_\pi+E_K)\,,\nl
V_{13,2}&=&-\frac{2g^2 \sqrt{m_{B_s}m_{B}}}{f_\pi^2}  \frac{m_{B}}{u-m_{B^{*}}^2} \vp_\pi\cdot \vp_K  \,,\nl
V_{14}&=&\frac{2ig^2\sqrt{m_{B_s} m_{B^{*}}}}{f_\pi^2} \frac{m_{B^*}}{u-m_{B^*}^2} \vep^{\prime *}\cdot (\vp_{K}\times \vp_{\pi})\, , \nl 
V_{23}&=& \frac{2ig^2\sqrt{m_{B_s^*} m_{B}}}{f_\pi^2} \frac{m_B}{u-m_{B^*}^2} \vep\cdot (\vp_{K}\times \vp_{\pi}) \,, \nl  
V_{24,1}&=&\frac{\sqrt{m_{B_s^{*}} m_{B^{*}}}}{2f_\pi^2}  (E_\pi+E_K) \vep \cdot \vep^{\prime *} \,,\nl
V_{24,2}&=&\frac{2g^2\sqrt{m_{B_s^*} m_{B^{*}}}}{f_\pi^2} \frac{m_{B^*}}{u-m_{B^*}^2} \nl
&\times& \left(\vep\cdot \vep^{\prime *}\,\vp_K \cdot \vp_\pi- \vep \cdot \vp_\pi\, \vep^{\prime *}\cdot \vp_K\right)\, .
\eea
In this equation the symbols $E_K$ and $E_\pi$ refer to the kaon and pion energy, in order, and 
 should be calculated accordingly to the scattering channels involved. 
E.g.  $V_{13}$ corresponds to  $B_s^0\pi^+ \to B^+ \overline K$ and then in the heavy-meson limit $E_\pi=\sqrt{s}-m_{B_s^0}$ and $E_K=\sqrt{s}-m_{B^+}$, 
with $s$ the standard Mandelstam variable. 
In terms of $E_K$ and $E_\pi$, the modulus of the kaon and pion three-momenta are  
 $|\vp_\pi|=\sqrt{E_\pi^2-m_\pi^2}$ and $|\vp_K|=\sqrt{E_K^2-m_K^2}$. The standard Mandelstam variable $u$ in the heavy-meson limit simplifies 
to $m_{B'}^2-2 m_{B'}E_\pi$, with $m_{B'}$ the mass of the final heavy meson.\footnote{In the study of $B\to\pi\pi l\bar \nu_l$ by Ref.~\cite{KangBl4}, 
 the full relativistic propagators are kept, rather than the leading term in the expansion of $1/m_P$ or $1/m_{P^*}$, as meant in Eq.~\eqref{190616.1},
in order not to miss any analytic structure of the $B^*\pi\pi$ triangle graph in $\pi\pi$ rescattering. 
However, in our current situation, we are dealing with two-point unitarity loop functions and
 the form of the heavy meson propagator in Eq.~\eqref{eq:potential}  can be used without problem.} 
In this form the propagators in Eq.~\eqref{eq:potential}, $m_{B'}/(u-m_{B^*}^2)$, tend to the standard form in HMChPT. E.g. for 
$V_{13}$
\begin{align}
\label{190616.1}
\frac{m_B}{u-m_{B^*}^2}\to -\frac{1}{2(\Delta+E_\pi)}~,
\end{align} 
where $\Delta$ is the mass difference $m_{B^*}-m_B$, and similar results apply to the other transitions between channels.  
Note also that factors of $\sqrt{m_P}$ and $\sqrt{m_{P^*}}$ have been included in the definition of $P$ and $P^*$ fields.

 The symbols $\vep$ and $\vep'$ in Eq.~\eqref{eq:potential}  represent the polarization vectors 
for the incoming and outgoing heavy vector mesons in spherical basis, respectively. 
As a result of the large masses of the $B$ and $B^*$  we can take the polarization vectors in the rest frame of the heavy vectors. 
 In the Cartesian basis they can be written as
\beq
\vec\epsilon_1=\begin{pmatrix} 1\\ 0\\ 0\end{pmatrix}\, , \quad \vec\epsilon_2=\begin{pmatrix} 0\\ 1\\ 0\end{pmatrix}\, ,\quad  \vec\epsilon_3=\begin{pmatrix} 0\\ 0\\ 1\end{pmatrix}\, .
\eeq
In the spherical basis they are given by 
\beq
\vep(\pm 1)=\mp \frac{1}{\sqrt{2}} (\vec \epsilon_1 \pm i \vec \epsilon_2)\, ,\quad  \vep(0) = \vec \epsilon_3,
\eeq
and take the form
\beq
\vep(\pm 1)=\mp \frac{1}{\sqrt{2}}  \begin{pmatrix} 1\\ \pm i\\ 0\end{pmatrix}\, , \quad \vep_3=\begin{pmatrix} 0\\ 0\\ 1\end{pmatrix}\, .
\eeq
 It follows then that both $V_{13,1}$ and $V_{24,1}$ do not involve any angular dependence, and thus they only contribute to $S$-wave.
 The $\sqrt{m_1m_2}$ factors appearing in Eq.~\eqref{eq:potential} can be safely replaced by just $m_{B_s^0}$, since the biggest difference can
be $(m_{B_s^*}-m_{B^+})/m_{B_s^0}\sim m_\pi/m_{B_s^0} < 3\%$. 
 However, the mass splittings play a more important role in the 
unitarity two-point loop functions due to the proximity of the associated branch-point singularity for the first and second channels to
 the $X(5568)$ nominal mass, i.e., the most important heavy-quark symmetry breaking corrections arise from the unitarization. 
Such strategy has also been widely adopted in many instances of coupled-channel studies, see e.g. the recent 
$\pi\Sigma_c$ scattering study in Ref.~\cite{Lambdac}.
In principle, one should work in the CM, however, we have checked that  in the kinematic region that we 
consider, $5.5\,\text{GeV} < \sqrt{s} < 5.7\,\text{GeV}$, differences are only about a  few percent compared with 
 Eq.~\eqref{eq:potential}, which is worked out in rest frames of the heavy mesons. 
For instance, the term $\sqrt{m_{B^+} m_{B_s^0}} (E_K+E_\pi)$ in $V_{13,1}$ comes from $\sqrt{m_{B^+} m_{B_s^0}} v\cdot(p_K + p_\pi)$, 
a term that is replaced by $(p_{B_s^0}+p_{B^+})\cdot (p_K + p_\pi)/2$ using the covariant Lagrangian \cite{Yan} 
(the one used in Ref.~\cite{Oset5568}).
It turns out that  these two forms differ only by a few percent.
 Specially, the ratio of the covariant one (that could be the most precise form) to the form used here  in the region around the mass of $X(5568)$
 is inside the range $[0.998,\,1]$. 
 The lack of relevance of these corrections
 is due to the small velocity involved in the Lorentz transformations from the heavy-meson rest frames to the  CM one.

 We proceed now to the $P$-wave projection of the tree-level amplitudes in Eq.~\eqref{eq:potential}.
 To that end, there are several ways to perform the partial wave decomposition. 
One consists of working first in the helicity basis \cite{Jacob}, and then transforming into partial wave $lSJ$ basis,
 see e.g., \cite{NN-helicity-PW} for the case of nucleon-nucleon ($NN$) scattering. 
Another one employs the tensor formalism \cite{Chung}, see e.g., \cite{VanKolck} again for  $NN$ scattering. 
We follow here the formalism developed in Ref.~\cite{Lacour}, which establishes that the partial-wave amplitude for the  scattering
 $1\,+\,2\to 1'\,+\,2'$ can be written generally as 
\begin{align}
\label{eq:PWD}
&V_{JI}(l'\,S',l\,S) = \frac{Y_l^0(\hat z)}{2J+1} \sum \langle s_1'\sigma_1's_2'\sigma_2'|S's_3'\rangle \langle s_1\sigma_1 s_2\sigma_2|Ss_3\rangle\nl
\times & \langle l 0 S s_3|J s_3 \rangle \langle l' m' S' s_3'|J s_3 \rangle  \langle \tau_1' \alpha_1' \tau_2' \alpha_2'| I i_3\rangle \langle \tau_1 \alpha_1 \tau_2 \alpha_2| I i_3\rangle \nl
\times & \int d\Omega\, V(\vec p\,', \sigma_1' \alpha_1' \sigma_2' \alpha_2'\, ; \,  p\hat z, \sigma_1 \alpha_1 \sigma_2 \alpha_2) Y_{l'}^{m'} (\theta, \phi)^*, 
\end{align}
where the repeated indices for the third components of spin-like quantities (including isospin) are summed.
 On the left hand side of the previous equation  $(l,\,S)\, [(l',\,S')]$ is the orbital angular momentum
 and total spin in the initial [final] state, and $I$ and $J$ are the total isospin and angular momentum, in order,
which are conserved in the scattering process; in our present case  $I=J=1$. 
 On the right hand side of Eq.~\eqref{eq:PWD}, we have denoted the Clebsch-Gordan coefficients (CGC) by 
 $\langle j_1 m_1 j_2 m_2| j m\rangle$, where $m_1$, $m_2$ and $m=m_1+m_2$ are the third components of the angular momenta $j_1, j_2, j$, respectively. 
 One can easily identify in Eq.~\eqref{eq:PWD} that the product of the two CGCs in the first line corresponds
 to the composition of the final/initial spins to the total one. Analogously, the product of the first two CGCs in the second line 
is for the composition of the orbital angular momentum and total spin to give the total angular momentum,
 and the last two CGCs in the same line proceed similarly for the isospin. 
 Within the integration the scattering amplitude, here referred by $V$, is calculated in the well-defined three-momentum basis 
with the corresponding isospin and spin indices  for each particle.
Besides, $Y_l^m(\theta,\phi)$ is the standard spherical harmonics function, and fixing the $z$-axis along the initial three-momentum 
$\vp$ the solid angle of the final three-momentum $\vp'$ is parameterized by $(\theta, \phi)$.
 For our current case, the spin and isospin parts just contribute with a  factor of one  each. 
In case of the scattering involving pure spin-0 particles, Eq.~\eqref{eq:PWD} simplifies into the well-known form
\beq
T(s,\cos\theta)=\sum_l (2l+1) T_l(s) P_l(\cos\theta),
\eeq
with $P_l(\cos \theta)$ denoting a Legendre polynomial.

The resulting $P$-wave projections of the amplitudes in Eq.~\eqref{eq:potential} as follows from Eq.~\eqref{eq:PWD} are  
\bea\label{eq:Pwave potential}
V_{13,\,l=l'=1}&=&-\frac{2g^2m_{B}\sqrt{m_{B_s}m_B}}{3f_\pi^2(u-m_{B^*}^2)} |\vp_\pi| |\vp_K|\, ,\nl
V_{14,\,l=l'=1}&=&-\frac{\sqrt{8}g^2 m_{B^*} \sqrt{m_{B_s}m_{B^*}}}{3f_\pi^2(u-m_{B^*}^2)} |\vp_\pi| |\vp_K|\, ,\nl
V_{23,\,l=l'=1}&=&-\frac{\sqrt{8}g^2 m_B\sqrt{m_{B_s^*}m_{B}}}{3f_\pi^2(u-m_{B^*}^2)} |\vp_\pi| |\vp_K|\, ,\nl
V_{24,\,l=l'=1}&=&\frac{2g^2 m_{B^*}\sqrt{m_{B_s^*}m_{B^*}}}{3f_\pi^2(u-m_{B^*}^2)} |\vp_\pi| |\vp_K|.
\eea

\section{Results and discussion}
\label{sec:results}

The potential calculated in Sec.~\ref{sec:potential} is the leading-order prediction in HMChPT. 
We want to take into account final state interactions by implementing the unitarization procedure
 of Ref.~\cite{NPA620}, where the potential is treated on shell and the Bethe-Salpeter
 equation can then be solved algebraically.
 This procedure can be also seen as a result of solving the $N/D$ method,
 firstly taking only into account the right-hand or unitarity cut and then including it perturbatively by 
matching with the chiral expansion order by order \cite{oller.110616.1,oller.110616.2,OllerMeissnerPLB}. 
In the dimensional regularization scheme, the renormalized unitarity two-point loop function 
for channel $i$  (with masses $m_{1i}$,\, $m_{2i}$) can be expressed as \cite{Hooft,gfunc2006} 
\begin{align}
\label{eq:Gdim}
G_i(s)=& \frac{1}{(4\pi)^2}\left( \alpha_i(\mu) + \log\frac{m_{2i}^2}{\mu^2}-\varkappa_+\log\frac{\varkappa_+-1}{\varkappa_+}\right.\nn\\
-&\left. \varkappa_-\log\frac{\varkappa_--1}{\varkappa_-}
\right)~,
\end{align}
where $\alpha_i(\mu)$ is a subtraction constant at the renormalization scale $\mu$ and 
 \begin{align}\label{eq:kappa}
\varkappa_\pm=&\frac{s+m_{1i}^2-m_{2i}^2}{2s}\pm \frac{k_i}{\sqrt{s}}~,\nn\\
k_i=&\frac{\sqrt{(s-(m_{1i}-m_{2i})^2)(s-(m_{1i}+m_{2i})^2)}}{2\sqrt{s}}~,
\end{align}
 The coupled-channel scattering $T$-matrix can be written as \cite{NPA620,oller.110616.1}\footnote{Note the change of sign in the convention 
for $T(s)$ and $V(s)$ used here with respect to that in Refs.~\cite{NPA620,oller.110616.1}.}
\beq\label{eq:T-matrix}
T^{-1}(s)=V^{-1}(s)+G(s).
\eeq
In our current case, $V(s)$ and $T(s)$ are $4\times 4$ matrices incorporating the coupled-channel transitions.
 The potential matrix $V(s)$ has been elaborated in detail in Sec.~\ref{sec:potential} and 
we will use its  $P$-wave projection, given in Eq.~\eqref{eq:Pwave potential}.
 The four-dimensional diagonal matrix $G(s)$  contains the two-point loop functions for all channels.
 If there is a pole at $s_X$, the determinant of the right hand of Eq.~\eqref{eq:T-matrix} should be zero at $s_X$.
 We denote by $\Delta(s)$ to this determinant, namely,
\beq\label{eq:determinant}
\Delta(s)=\text{det}\,[V^{-1}(s)+G(s)].
\eeq
Since a resonance appears in the unphysical Riemann sheets, we need the analytic continuation of $G_i(s)$ from the physical (first) sheet
 (i.e., Eq.~\eqref{eq:Gdim}) to the unphysical one associated to this channel. This analytical continuation is given by \cite{NPA620}
\beq\label{eq:cont}
G^{II}_i(s)=G_i(s)+i\frac{k_i}{4\pi\sqrt{s}},
\eeq
where $k_i$ is calculated such that $\text{Im}\, k_i>0$ as in the physical sheet.
 The experimentally reported mass of $X(5568)$ is around 5568 MeV, cf.~Eq.~\eqref{eq:pole position},
which stays between the mass thresholds of channels  $B_s^*\pi^+$ (2) and  $B^+\overline K^0$ (3), i.e., $m_{B_s^{\prime 0}}+m_\pi< m_X < m_B+m_K$. 
Thus, we work in the unphysical Riemann sheets of channels (1) and (2), and in the physical sheets of channels (3) and (4), because 
the Laurent expansion around the resonance pole embraces the physical energy axis between channels (2) and (3). 
This Riemann sheet is named as (1,1,0,0), where the number 1 indicates that the analytical continuation to the unphysical sheet 
of the corresponding channel is applied.
 The $G$ matrix in the (1,1,0,0) Riemann sheet is  
\beq
G(s)=\text{diag}\,\{G_1^{II}(s),\, G_2^{II}(s),\, G_3(s),\, G_4(s)\}
\eeq
 Other sheets in this notation can be inferred easily (and indeed they have been also explored, particularly the closer (1,0,0,0) sheet).

\begin{figure}[h]
\begin{center}
\includegraphics[angle=0, width=.5\textwidth]{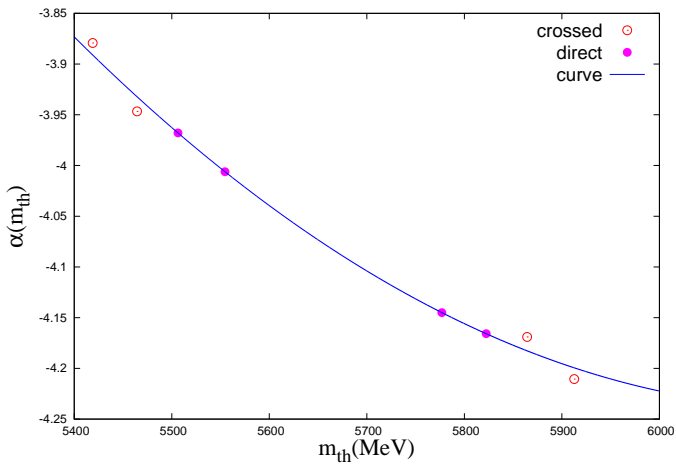}
\end{center}
\caption[pilf]{\protect \small Subtraction constants $\widetilde{\alpha}_i(M_\rho)$ and $\alpha_i(M_\rho)$ as a function of the threshold mass, 
given by the empty and filled circles, in order. 
 The solid line is a parabola fit to the former ones, cf. Eq.~\eqref{170616.1}. 
\label{fig:120616.1}
}
\end{figure}

We now elaborate about the subtraction constant in Eq.~\eqref{eq:Gdim}.
The crossed $u$-channel $P$-wave  scattering settles the constraint that after unitarizing the exchange of the $B^{*0}$
 must give rise to a simple pole 
located at the same position as in the Born term, that is, at $u=m_{B^{*0}}^2$.
 The channels involved in the $s$-channel  scattering becomes after $u$-crossing  
 $B_s^0 K^0$ (1), $B_s^* K^0$ (2),  $B^+ \pi^-$ (3) and  $B^{*+}  \pi^-$ (4).  
The $P$-wave projected leading-order HMChPT amplitudes in the $u$-channel look the same as those in Eq.~\eqref{eq:Pwave potential}. 
However, now $u$ does not vary between channels (in the same was as $s$ is the same for all channels in $s$-channel scattering),  
the three-momenta are calculated in terms of $u$ (not $s$) and one has to keep in mind the rearrangement of mesons in the new
 channels with zero strangeness. 
In the following, in order to avoid using the same symbols for both $s$- and $u$-channel scattering we place a tilde on top of the symbols for the latter case.  
 The unitarization process is then performed making use of Eq.~\eqref{eq:T-matrix} and, in order that 
the resulting partial-wave amplitudes have a simple pole at $u=m_{B^{*0}}^2$, it is required that all the two-point unitarity wave functions vanish at that point. 
This is clear because then  $\widetilde{T}(u)=[I+\widetilde{V}(u)\cdot \widetilde{G}(u)]^{-1}\cdot \widetilde{V}(u)$ (a rewriting of Eq.~\eqref{eq:T-matrix}) 
 and the condition $\widetilde{G}(m_{B^*}^2)=0$ imply that the matrix $[I+\widetilde{V}(u)\cdot \widetilde{G}(u)]^{-1}$ does not contain
 any pole at $u=m_{B^{*0}}^2$, so that both $\widetilde{T}(u)$ and $\widetilde{V}(u)$ have a simple pole there.\footnote{Another possible solution is to 
employ in the Born terms a bare mass for the particle exchanged that is dressed so as to end with the proper pole, see e.g. Ref.~\cite{oset.210616.3}. We 
have followed here the procedure of Refs.~\cite{alarcon.210616.1} and \cite{alarcon.210616.2}, which is the most suitable one when calculating higher orders 
 in the chiral perturbative series. This is so because by using the physical mass in the Born terms
 one can avoid  double-pole terms that otherwise would appear in the perturbative series, see also Ref.~\cite{becher.210616.4}.} 
The subtraction constants for the $u$-channel states, indicated by $\widetilde{\alpha}_i$, can then be evaluated straightforwardly from the condition 
$\widetilde{G}(m_{B^{*0}}^2)=0$, cf. Eq.~\eqref{eq:Gdim}, and the resulting values for $\mu=770~{\text{MeV}}\simeq M_\rho$ 
 are plotted by the empty circles  in Fig.~\ref{fig:120616.1} as a function of the two-body threshold.
 These values are around $-4$ and  we observe that they mainly follow a linear behavior with some small curvature.
 We then fit these points with the curve 
\begin{align}
\label{170616.1}
\widetilde{\alpha}(m_{\rm th})=a+b m_{\rm th}+ c m_{\rm th}^2~, 
\end{align}
which reproduces rather well the input points, as shown by the solid line in Fig.~\ref{fig:120616.1}.  Once this is set up we can reliably determine 
from Eq.~\eqref{170616.1} the value of the $s$-channel subtractions constants $\alpha_i(\mu)$ for our original problem on  $s$-channel $P$-wave scattering.
 The values are shown in Fig.~\ref{fig:120616.1} by the filled circles. 
We then search for the zeros of $\Delta(s)$, cf. Eq.~\eqref{eq:determinant},  
in the complex plane of (1,1,0,0) and (1,0,0,0) sheets. The result of this search is that no pole is found there 
in the region of the $X(5568)$.

The values for the subtraction constants in Fig.~\ref{fig:120616.1} are indeed quite close to their natural values according to Ref.~\cite{OllerMeissnerPLB}. 
This reference exploits the fact that an apparent meaning for ``natural'' can be found in three-momentum cutoff regularization. Namely, 
 the three-momentum cutoff $\Lambda$ should be around the chiral expansion scale, or more intuitively speaking, 
 it should correspond to the scale in configuration space around the finite size of particles. 
Here the natural value for $\Lambda$ is about the $\rho$ mass or 1 GeV.
 By comparing Eq.~\eqref{eq:Gdim} with the one in the cutoff regularization (Eq.~\eqref{eq:Gcut} below), one finds \cite{OllerMeissnerPLB}
\beq\label{eq:natural}
\alpha_i(\mu_i)=-2\log\left(1+\sqrt{1+\frac{m_{2i}^2}{\mu_i^2}}\right),
\eeq
where in the channel $i$, $m_{2i}$ is  the heavier mass, e.g., for channel  $B_s^0\pi^+$, $m_{21}=m_{B_s^0}$ and $m_{11}=m_{\pi^+}$.
Powers  of $k_i^2$ and $m_{1i}/m_{2i}$ have been neglected in obtaining Eq.~\eqref{eq:natural}.
As with the values of the subtraction constants given by the filled points in Fig.~\ref{fig:120616.1}, 
the search for zeros of $\Delta(s)$ using Eq.~\eqref{eq:natural}  is also fruitless.

A more radical strategy to try to reproduce the $X(5568)$ pole would be to treat the subtraction constants $\alpha_i$ as free parameters, so as  
to impose a pole at $\sqrt{s_X}\approx 5568 \,-\, i\, 11$ MeV. Of course, in this form the $u$-channel crossed-exchange constraint $\widetilde{G}(m_{B^{*0}}^2)=0$ 
is not satisfied.  
Using a unique subtraction constant shared by all the four loop functions, one can solve the equation $|\Delta(s_X)|=0$ and  
four sets of {\it complex} solutions result then, which are not acceptable because they drive to violations of unitarity 
(on top of the mentioned one from crossing).
When we fix two {\it real} subtraction constants around their values in Fig.~\ref{fig:120616.1} in a short region,
 one can determine the other two {\it real} subtraction constants by solving the equations $\text{Re}\, \Delta(s_X)=0$ and  $\text{Im}\, \Delta(s_X)=0$.
 It turns out that no solutions in the natural-size range can be found; e.g., by fixing 
$\alpha_2=-4,\,\alpha_4=-4.15$, one gets $\alpha_1=-1.8,\, \alpha_3=-114.8$ or $\alpha_1=382.5,\, \alpha_3=-1.2$.
 There are other regions in the parametric space $(\alpha_2,\alpha_4)$ that also drive to real $(\alpha_1,\alpha_3)$ parameters,
 but all of them are very far from the region of natural values for the $\alpha_i(\mu)$.
 Thus, there is no appropriate solution that can give rise to an $X(5568)$ dynamically generated from 
the interactions of $B_s\pi^+$ plus coupled-channels in $P$-wave scattering. Of course, as follows from the discussion, 
one could impose the $X(5568)$ pole but then it would correspond to a new  elementary degree of freedom of unknown origin. 
That is, the generation of the pole could be reproduced but we could not deduce any explanation about its origin, as we are seeking.

Next, we take the functions $G_i(s)$ calculated with a common three-momentum cutoff, $\Lambda$,
to further confirm the above findings and because it is also used in the study  of Ref.~\cite{Oset5568} 
on $B_s\pi$ and $B\overline{K}$  $S$-wave scattering.
 The two-point unitarity loop function regularized with a three-momentum cutoff 
$\Lambda$ was calculated in Ref.~\cite{OllerPelaez}. It reads
\bea \label{eq:Gcut}
G_i(s)&=&\frac{1}{16\pi^2}\left[-\frac{\delta}{2s}\log\frac{m_{1i}^2}{m_{2i}^2}+\frac{\nu_i}{2s}\left(\log\frac{s-\delta+\nu_i\lambda_1}{-s+\delta+\nu_i\lambda_1}\right.\right.\nl
&& +\left.\log\frac{s+\delta+\nu_i\lambda_2}{-s-\delta+\nu_i\lambda_2}\right)+\frac{\delta}{s}\log\frac{1+\lambda_1}{1+\lambda_2}\nl
&&\left.\vphantom{\frac{1}{16\pi^2}}-\log[(1+\lambda_1)(1+\lambda_2)]+\log\frac{m_{1i}m_{2i}}{\Lambda_i^2}\right],
\eea 
with 
\bea
\nu_i&=&2\sqrt{s}k_i\, ,\nl
\delta&=&m_{2i}^2-m_{1i}^2\, ,\nl
\lambda_1&=&\sqrt{1+\frac{m_{1i}^2}{\Lambda_i^2}}\,,\quad \lambda_2=\sqrt{1+\frac{m_{2i}^2}{\Lambda_i^2}}\,.
\eea
and $k_i$ is given in Eq.~\eqref{eq:kappa}. Both Eq.~\eqref{eq:Gdim} and Eq.~\eqref{eq:Gcut} are symmetric under the exchange of $m_{1i}\leftrightarrow m_{2i}$.
The analytic continuation to the unphysical sheet is also given by Eq.~\eqref{eq:cont}.
 We have looked for poles of $1/\Delta(s_X)$ by varying  $\Lambda$ (shared by all the four $G_i(s)$ in Eq.\eqref{eq:Gcut}),
and find that this function changes smoothly in a rather 
large region of $\Lambda$ (a much wider region than the natural one around 1~GeV). 
 Being inspired by the functional dependence of the estimation of the natural value for the subtraction constants on $m_{2i}$, cf. Eq.~\eqref{eq:natural},
 we further assume that channels (1) and (2) share a cutoff $\Lambda_1$, and channels (3) and (4) do so with a cutoff $\Lambda_2$.
 We make a three-dimensional plot by varying these two cutoffs and, again, the result is negative.

Generally speaking, the $B_s\pi$ and $B\overline{K}$ $S$-wave contributions calculated  in Ref.~\cite{Oset5568} are around five times larger than any individual
$P$-wave piece in the region of the $X(5568)$ resonance.
 When we replace $V(s)$ in Eq.~\eqref{eq:determinant} by $15\times V(s)$, we then find a pole at $5789\, -\, i\, 21$ MeV in the (1,1,1,0)
 sheet (which is in between the $B\overline K$ and $B^*\overline K$ mass thresholds) with a cutoff in the natural range, $\Lambda=1$ GeV. 
  This fact clearly indicates that the leading-order HMChPT $P$-wave potential that we have calculated is too weak to become resonant 
with a natural cutoff applied to the two-point unitarity loop functions. 
To a lesser extent this conclusion also holds for the  $S$-wave case, as worked out in Ref.~\cite{Oset5568},
whose results we also reproduce here.
 As for the dimensional regularization form of $G_i(s)$, cf. Eq.~\eqref{eq:Gdim}, 
the authors of Ref.~\cite{Oset5568} obtain $\alpha(\hat{\mu})=-0.97\pm 0.02$,  with $\hat{\mu}=(m_{B_s^0}+m_{B^{*+}})/2=5323$~MeV. When 
translated to our scale this would correspond to $\alpha(M_\rho)\simeq -4.9$, which is clearly out of the range 
of natural values shown in Fig.~\ref{fig:120616.1}. Indeed, we cannot generate such a value of $\alpha$ making use of Eq.~\eqref{170616.1} because 
the curvature would become dominant for higher mass thresholds and $\alpha$ does not become smaller than around $-4.3$. If we pursue along with the 
approximate linear
 behavior in Fig.~\ref{fig:120616.1}, it would imply a mass threshold clearly larger than 7000~MeV, more than 1200 MeV higher than the $B\overline{K}$ threshold. 
 However, if the isovector $B_s\pi-B\overline{K}$  $S$-wave potential is multiplied by a factor of 3 we then find a pole at $5548\,-\,i\,23$ MeV
 with $\Lambda =1$~GeV.  This pole does not quantitatively agree with the experimental values for the mass and width of the $X(5568)$, but our point is
to show that more strength in the potential is needed to generate the $X(5568)$  with natural cutoffs or natural values for the subtraction constants. 
  We remark here that these exercises are just done for pure academic purposes, and that there is no physical reason to enhance the potential calculated  
by such factors.
  However, it helps us to understand better that if the observed $X(5568)$ is not found in the scattering amplitude evaluated with a natural cutoff
  is because the weakness of the leading-order HMChPT transition potentials, agreeing with the inference from Burns and Swanson \cite{Swanson}. 
These findings point to the direction that is hard to explain theoretically the experimentally reported $X(5568)$ by the D0 Collaboration. 
 At least it cannot be described in a straightforward manner as a dynamically generated resonance by coupled-channel scattering, 
 once the results of the $S$-wave study of Ref.~\cite{Oset5568} and the current $P$-wave attempt are taken simultaneously.

By an obvious replacement of  the relevant masses, one can also explore the charm sector, namely, 
we have at hand the scattering of $D_s^+\pi^0$,\, $D_s^{*+}\pi^0$,\, $D^+\overline K^0$ and $D^{*+}\overline K^0$, 
with thresholds 2103.44, 2247.27, 2362.6 and 2499.97 MeV in order. 
No resonant pole for a partner state of the $X(5568)$ 
is found when a three-momentum cutoff around 1 GeV is used, in the lines of the criticism of Ref.~\cite{FKGuo}.

\section{Summary and conclusions}
\label{sec:conclusion}

The recently observed exotic hadron candidate $X(5568)$ by the D0 Collaboration \cite{D0} in the $B_s^0\pi^\pm$ event distribution 
with quantum numbers $0^+/1^+$ has triggered interesting theoretical and experimental questions.   
In the present work, we first perform the analysis of the experimental event-distribution data by using a $P$-wave Breit-Wigner parameterization 
for the resonance signal function, instead of an $S$-wave BW as used by the D0 Collaboration in its analysis \cite{D0}. 
We fix the mass and  width of the $P$-wave BW to the values obtained in Ref.~\cite{D0} and
 the resulting fit reproduces data quite closely, being both fits (either assuming $S$- or $P$-wave $B_s^0\pi^\pm$) of similar quality. 

 We then make a complementary analysis by considering the  $P$-wave coupled-channel scattering between the channels 
$B_s^0\pi^+$, $B_s^*\pi^+$, $B^+\overline K^0$ and $B^{*+}\overline K^0$. Notice that then $B_s\pi$ and $B_s^*\pi$ can couple each other and, 
 indeed, the mass threshold for $B_s^*\pi$ is much closer to the nominal mass of the $\X$ than the one for $B_s\pi$. This is an 
interesting point when one is considering the possible ``molecular'' nature for this resonance. It is clear that the $B_s\pi$ 
and $B_s^*\pi$ three-momenta are small (which presumably makes a $P$-wave scattering small),
 but the off-diagonal transitions involving the heavier $B\overline{K}$ and $B^*\overline{K}$ states have much larger {\it unphysical} three-momenta, 
determined by analytical continuation. 
The transition potential matrix $V(s)$ is calculated in  leading-order Heavy Meson Chiral Perturbation Theory and it is then 
unitarized employing techniques of Unitary Chiral Perturbation Theory \cite{NPA620,oller.110616.1,OllerMeissnerPLB}. 
 The subtraction constants appearing in the unitarization process are determined with almost coincident values, either by requiring
 the reproduction of the $B^{*0}$ simple pole in the $u$-channel $P$-wave scattering or by employing its natural values. 
 We then search for poles in the resulting $T$-matrix  and no pole that could be associated to the  $X(5568)$ is found.   
 The same conclusion is obtained when using the unitarity two-point loop functions calculated 
in terms of a  three-momentum cutoff varied within a wide 
range around 1~GeV. 
 If we further release the subtraction constants the $\X$ pole could be imposed, but the resulting values for the subtraction constants 
prevent us from consider such a pole as dynamically generated from the $P$-wave interactions 
between the aforementioned two-body channels.  In other terms, no explanation for its nature would then be obtained within our approach.  
  Replacing the heavy $B$-like mesons by the corresponding charm ones, the charm sector can also be explored  and similar results arise as well.
 We conclude that the main reason for the absence of poles is the weakness of the interactions.

In summary, by combining our  $P$-wave study  in the current work with the recent $S$-wave study of Ref.~\cite{Oset5568},  
 the new state $X(5568)$ observed by D0 the Collaboration cannot be likely interpreted as a ``molecular'' resonance generated 
dynamically from two-body coupled-channel scattering.
This point has also been supported by  several other groups, see e.g. \cite{Swanson,FKGuo,molecular}.
The nature of $X(5568)$ is still unsettled, and even its existence is inconclusive
 recalling also the experimental non-observation of this state by LHCb \cite{LHCb}.  
More theoretical and experimental efforts are still needed in order to settle these issues.

\section*{Acknowledgements}
We would like to thank E.~Oset for a careful reading of the manuscript and interesting comments and 
 X.W.K. would like to thank also M.~Albaladejo for a clarification on the potential used in Ref.~\cite{Oset5568}.
This work is supported in part by the MINECO (Spain) and ERDF (European Commission) grant FPA2013-40483-P and the Spanish
Excellence Network on Hadronic Physics FIS2014-57026-REDT.


\end{document}